%% file: main.tex
\documentclass[aps,prd,twocolumn,superscriptaddress,groupedaddress]{revtex4-2}

\input{preamble}

\usepackage{xcolor}

\newcommand{\ellbar}{\hbox{\it \l}\,}

\usepackage{subfiles}
\usepackage{cancel}

\begin{document}

\title{Cosmological constraints from harmonic space analysis of DES Y3 3x2 clustering}
\author{Utkarsh Giri{\href{https://orcid.org/0000-0001-5553-9167}{\includegraphics[height=8pt]{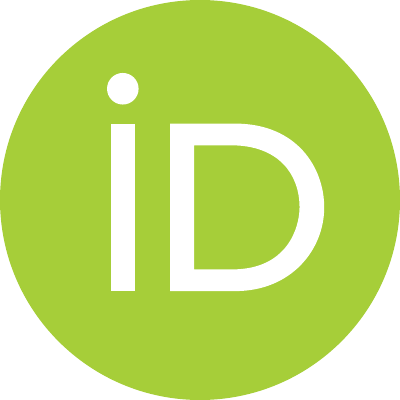}}}}\email[Correspondence email address: ]{ugiri@wisc.edu}
    \affiliation{Department of Physics, University of Wisconsin-Madison, Madison, WI 53706, USA}
\author{Sai Chaitanya Tadepalli{\href{https://orcid.org/0000-0001-9947-4748}{\includegraphics[height=8pt]{plots/ORCIDiD_iconvector.pdf}}}}\email[Email address: ]{stadepalli@wisc.edu}
    \affiliation{Department of Physics, University of Wisconsin-Madison, Madison, WI 53706, USA}
\date{\today}

\begin{abstract}
The large-scale distribution of matter, as mapped by photometric surveys like the Dark Energy Survey (DES), serves as a powerful probe into cosmology. It is especially sensitive to both the amplitude of matter clustering ($\sigma_8$) and the total matter density ($\Omega_m$). The fiducial analysis of the two-point clustering statistics of these surveys is invariably done in configuration space where complex masking scheme is easier to handle. However, such an analysis inherently mixes different scales together, requiring special care in modeling. In this study, we present an analysis of \textsc{DES Y3} 3x2 clustering data in harmonic space where small and large scales are better separated and can be neatly modeled using perturbative techniques. Using conservative scale cuts together with Limber approximation and a Gaussian covariance assumption in a first study, we model the clustering data under a linear bias model for galaxies, incorporating comprehensive treatment for astrophysical effects. We subsequently extend this fiducial analysis to explore a third-order biasing prescription. For our fiducial analysis, we get $S_8=0.789\pm0.020$, consistent with the configuration space analysis presented by the DES collaboration, although under our different modelling choices, we find a preference for a lower $\Omega_m$ and a higher $\sigma_8$. The analysis sets the stage for a future search for signatures of primordial non-Gaussianity and blue-tilted isocurvature perturbations from photometric surveys.
\end{abstract}
\maketitle

\section{Introduction}
Over the last few decades, our knowledge and understanding of the Universe have grown by leaps and bounds, thanks mainly to massive amounts of high-quality data accumulated by Cosmic Microwave Background (CMB) experiments like WMAP \cite{hinshaw2012nineyear} and Planck \cite{Planck:2018vyg} and spectroscopic and photometric galaxy surveys like the SDSS\cite{2019BAAS...51g.274K} and DES \cite{dark2016dark} experiments. This has led to the emergence of a very successful and \emph{mostly} concordant model -- the $\Lambda$CDM model, in which the universe is dominated by dark energy in the form of a cosmological constant with structures forming in potential wells sourced by cold dark matter \citep{Peebles:2002gy, Bull:2015stt}. 

Of late, however, several tensions have started to appear in the $\Lambda$CDM model and have grown in significance over time as the precision of experiments and analysis has advanced \citep{Perivolaropoulos:2021jda, Bull:2015stt}. Among these, the Hubble tension \cite{DiValentino:2020zio} and the $S_8$ tension \cite{DiValentino:2020vvd} are arguably the two most prominent ones. Both the Hubble tension as well as the $S_8$ tension pertain to a significant discrepancy between derived values of the parameters from low and high redshift observables. The $S_8$ tension, in particular, refers to the tentative evidence for a slightly lower value of $S_8$ from galaxy survey experiments when compared to its value derived from CMB experiments. Although a  clear resolution has not yet been found, these recent developments have shown the importance of performing comparative analysis between different datasets and using different analysis approaches. 


In this study, we present a harmonic space analysis of the 3x2 clustering statistics of DES Y3 data using the pseudo-$C_l$ framework. The pseudo-$C_l$ estimator is a well-tested and state-of-the-art approach for analyzing two-point statistics in harmonic space. We validate our setup by first analyzing the DES Y3 cosmic shear angular power spectrum and reproducing the results of \citet{doux2022dark}. We subsequently analyze galaxy clustering, galaxy-galaxy lensing and cosmic shear clustering data - the so called 3x2 clustering statistcs, in a combined setup under the 6-parameter $\Lambda$CDM mode. 
For our fiducial 3x2 analysis, we use a conservative scale cuts on large and small scales with a small-scale cut of $k_{\max}=0.1~ \mathrm{Mpc^{-1}}$ for galaxies for all tomographic bins. In this first study, we work under the Limber approximation and use Gaussian covariance model for our dataset but use a comprehensive modeling for astrophysical and other systematic effects. Finally, we extend our fiducial analysis to explore a third-order biasing prescription. Our results are \emph{broadly} consistent with the DES Y3 configuration space analysis with an excellent agreement for the $S_8$ parameter. Notably, we find a preference for relatively lower $\Omega_m$ and a higher $\sigma_8$ compared to DES Y3 results which possibly arises due our different modeling choices. We leave systemic exploration of the modelling differences for a later work and briefly discuss them in \S\ref{sec:discuss}

After describing our dataset and the underlying theory in \S\ref{sec:dataset} and \S\ref{sec:theory} respectively,  we present our data processing in \S\ref{sec:processing} and our modeling in \S\ref{sec:modelling}. The inference framework is described in \S\ref{sec:inference}. Finally, in  \S\ref{sec:results} we present our results before concluding in \S\ref{sec:discuss}.

\vspace*{0.5cm}

\section{Dataset}

\label{sec:dataset}
Over six years from 2013 to 2019, the Dark Energy Survey mapped approximately 5000 square degrees of the southern sky in the $grizY$ band, using the 4m Blanco telescope at Cerro Tololo Inter-American Observatory in Chile. Several end-use catalogues have been publicly released by the collaboration from one and three years of data collection. More details of survey and the associated data products can be found in \citep{DES:2021wwk} and references therein.

The dataset used in this work comprises of a sample of 10.7 million lens galaxies from the \maglim~ galaxy catalogue and a sample of 100 million source galaxies in the calibrated shape catalogue released by the DES collaboration as part of their Y3 release \footnote{\url{https://des.ncsa.illinois.edu/releases/y3a2/Y3key-catalogs}}.

The \maglim~ lens catalogue is a magnitude limited sample of galaxies obtained by applying a magnitude cut in the $i$ band given by $i < 4z_{phot} + 18$. A further lower magnitude cut of $i < 17.5$ is applied to remove contamination from stars and other bright objects. This selection criteria was derived in \cite{DES:2020ajx} with the goal to optimize for the cosmological constraining power and results in a catalogue which has a decent photometric redshift uncertainties while at the same time having a very high number density. The \maglim~ catalogue was the first magnitude limited photometric catalogue used for constraining cosmological parameters. Each galaxy in the catalogue comes with an associated weight corresponding to the inverse of the estimated angular selection function. The redshifts of lens galaxies are estimated using the directional Neighbourhood Fitting (DNF) algorithm  \cite{DES:2020aks} and the entire sample is divided into 6 tomographic bins from $z=0.2$ to $z=1.05$ with bin edges $z = (0.20, 0.40, 0.55, 0.70, 0.85, 0.95, 1.05)$. The real-space analysis by \citep{DES:2021wwk} found issues with the last two bins of the lens sample and did not include them in their fiducial analysis. Following on their footsteps, we also discard the last two bins of the \maglim~ lens sample. Thus our data comprises of all four tomographic bins of the source catalogue and only the first four bins of the \texttt{Maglim} lens catalogue. 

The source galaxy catalogue consists of 100 million samples with their shapes estimated by the self-calibrating \textsc{Metacalibration} algorithm \citep{huff2017metacalibration, sheldon2017practical} using information from $riz$ bands. The \metacal~ algorithm is an approach for producing unbiased estimate of shear from observed galaxy ellipticity. It involves first applying small amount of synthetic shear to a de-convolved image of an observed source, before re-convolving it with a Point space function (PSF) model and estimating the response $R$ which linearly relates the observed shape to the true shear, bypassing the need for simulation based calibration. 
The entire source galaxy catalogue is divided into fours tomographic redshift bins of nearly equal number density, with redshifts inferred using a self-organizing map (SOM) algorithm.

\begin{figure*}[ht!]
    \centering
    \subfigure{\includegraphics[width=0.48\textwidth]{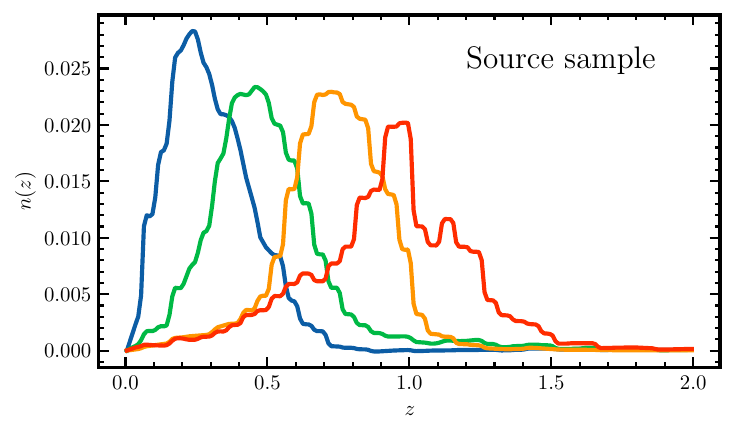}}
    \subfigure{\includegraphics[width=0.48\textwidth]{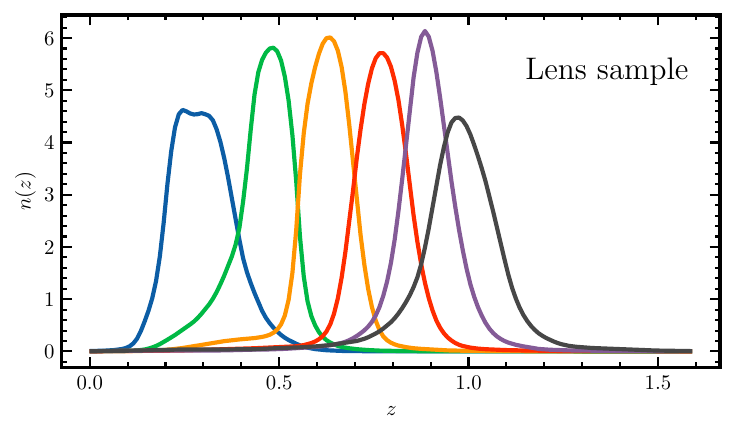}}
    \caption{Normalized photometric Redshift distribution $n(z)$ for the \metacal~source in four tomographic bins and for the \maglim~lens sample in six tomographic bins.}
    \label{fig:nz}
\end{figure*}

In Figure \ref{fig:nz}, we show the redshift distribution of the source and lens samples.

\subfile{sai_file_1_theory.tex}

\subsection{Cosmic Shear}
The lensing potential $\psi$ at a given position ($\vec{\theta}$) in the sky is a projection of 3D Newtonian potential $\Phi$
\begin{equation}
    \psi(\vec{\theta}) = 2 \int \frac{d\chi}{\chi} \Phi(\chi, \vec{\theta}\chi) q(\chi)
\end{equation}
where $q$ is called the lensing efficiency
\begin{equation}
    q(\chi) = \int d\chi' n(\chi') \frac{\chi' - \chi}{\chi}
\end{equation}
The lensing efficiency $q$ encodes information about the geometry of the Universe. The 3D potential is related to the matter field via the Poisson equation. The second orderr derivatives of the lensing potential defines the shear $\Gamma$ and convergence $\kappa$
The shear field $\gamma = \gamma_1 + i\gamma_2$ is a spin 2 field and is related to the potential via
\begin{equation*}
    \kappa = \frac{1}{4}(\eth \bar{\eth} + \bar{\eth} \eth )\psi(\vec{\theta}); \ \ \  \ \ \ \gamma(\vec{\theta}) = \frac{1}{2} \eth \eth \psi(\vec{\theta}) \nonumber \\
\end{equation*}
\begin{align}
    \kappa &= (\psi_{11} + \psi_{22})/2 \nonumber \\
    (\gamma_1, \gamma_2) &= \big((\psi_{11} - \psi_{22})/2, (\psi_{12} + \psi_{21})/2 \big)
\end{align}
In this paper, we work in the harmonic space where the shear $\gamma$ can be equivalently expressed in terms of spin-weighted spherical harmonics basis $_sY_{lm}$
\begin{equation}
    \gamma(\theta) = -\sum_{lm} \phantom{}_{\pm2}Y_{lm}(E_{lm} \pm iB_{lm})
\end{equation}
where $E$ and $B$ are the curl-free and divergence-free modes. 
The gravitational induced $E$-mode power spectrum is related to convergence power spectrum $C_{\kappa\kappa}$ as
\begin{equation}
    C_{EE}(l) = G(l)^2 C_{\kappa\kappa}(l)
\end{equation}
where $G_l$ is an $l$-dependent spin-prefactor given by
\begin{equation}
    G(l) = \frac{\ellbar}{(l+1/2)^2} = \frac{1}{(l+1/2)^2}\sqrt{\frac{(l+2)!}{(l-2)!}}
\end{equation}
for the spin-2 shear field. The prefactor is $\sim 1$ for $l\sim\mathcal{O}(10)$. Beyond this gravitational signal, effects like intrinsic alignment contribute additional power to the observed E-mode power spectrum. We use a non-linear alignment (NLA) \cite{2004PhRvD..70f3526H, Bridle_2007} model to model the intrinsic alignment.
\vspace{0.25cm}

To evaluate $C_{EE}^{ij}$ and $C_{gg}^{ij}$ (and cross-spectra $C_{gE}^{ij}$) we work under the Limber approximation \cite{limber, loverde2008extended} which is sufficiently accurate for $l>40$. \commentout{The 3x2 two-point statistics includes galaxy-galaxy, galaxy-galaxy lensing and lensing power spectra.} Under this approximation, the auto and cross power spectrum $C_{AB}^{ij}(l)$ for two tracers $A$ and $B$ for tomographic bins $i$ and $j$ can be written as \cite{limber, loverde2008extended}
\begin{equation*}
    C^{ij}_{AB}(l) = \int \frac{d\chi}{\chi^2} W_A^i(\chi) W_B^j(\chi) P_{mm}\bigg(k=\frac{l+1/2}{\chi},\ z(\chi)\bigg)
    \label{limber_projection}
\end{equation*}
where $P_{mm}$ is the matter power spectrum and $W_{A,B}^{i,j}$ is a kernel encoding the weight specific to a tracer $A$ for tomographic bin $i$. \commentout{For galaxy number counts, this is given by the galaxy number density i.e.\ $W_g^{i}(\chi)=n^i_g(z(\chi)$.} 

For the $E$-mode, the kernel has a slightly complex form given by
\begin{equation}
    W_E^i(\chi) \equiv G_l \frac{3}{2}H_0^2\Omega_m \frac{\chi}{a(\chi)}\int dz' n_s^i(z') \bigg[\frac{\chi(z')-\chi}{\chi(z')}\bigg]
\end{equation}
where $n_s^i$ is the source galaxy number density for bin $i$. Although our discussion is in terms of auto spectrum, it is straightforwardly generalized to get expressions for cross spectra as well.  

\section{Map making \& $C_l$ estimation}
\label{sec:processing}
\subsection{Map making}
\label{sec:map_making}
Before generating shear maps from source catalogue, the DES Y3 source catalogue requires to be corrected for  possible multiplicative or additive biases as outlined in \cite{doux2022dark}. For each redshift bin, we compute the weighted mean ellipticity and subtract that from the observed ellipticities of each galaxy. The \textsc{metacalibration} algorithm which self-calibrates the shear statistics, artificially shears the galaxies by a fixed amount. The resulting change in ellipticity is used to calibrate a \emph{total} shear response R which is used to normalize the measurement \cite{DES:2021bvc, DES:2021vln}
\begin{equation}
    e_i \rightarrow \frac{e_i - \langle e_i \rangle}{R}
\end{equation}
The de-trended catalogue is then used to generate weighted map of the tangential shear field $\gamma = (\gamma_1, \gamma_2)$ from observed ellipticity $e=(e_1, e_2)$ on a \texttt{healpix} \cite{Zonca2019, 2005ApJ...622..759G} grid with $N_{side}=4096$ for each tomographic bin \cite{nicola2021cosmic}. 
\begin{equation}
    \hat{\gamma}(\theta_p) = (\hat{\gamma}_1, \hat{\gamma}_2) = \bigg(\frac{\sum_{i \in p} w^s_i e_{1,i}}{\sum_{i \in p} w^s_i}, \frac{\sum_{i \in p} w^s_i e_{2,i}}{\sum_{i \in p} w^s_i}\bigg)
\end{equation}
where the sum is over source galaxies $i$ in the pixel $p$. $w^s_i$ is the weight assigned to that galaxy with and $(e_{1,i}, e_{2,i})$ is its measured ellipticity. The corresponding anisotropic noise variance map is obtained by \cite{nicola2021cosmic, doux2022dark}
\begin{equation}
    \sigma^{\gamma}(\theta_p) = \frac{\sum_{i \in p} {w^{s_i}}^2 (e_{1,i}^2 + e_{2,i}^2)}{(\sum_{i \in p} {w^s_i})^2}
\end{equation}
The above map-making operation is performed for each redshift bin separately and results in four shear maps corresponding to four tomographic bins of the shape catalogue.

\vspace{0.5cm}

For the \texttt{maglim} lens catalogue, the weighted galaxies counts are similarly deposited on a \texttt{healpix} map of $N_{side}=4096$. We then subtract the mean number count to get the galaxy overdensity map for each tomographic bin \cite{hadzhiyska2021hefty}.

\begin{equation}
   \delta_g(\theta_p)  = \frac{\sum_{i \in p} w^l_i}{\langle {\sum_{i \in p} w^l_i} \rangle} - 1
\end{equation}
where $w^l_i$ is the weight for lens galaxy $i$ in pixel $p$

\begin{figure*}[ht!]
    \includegraphics[width=0.98\textwidth]{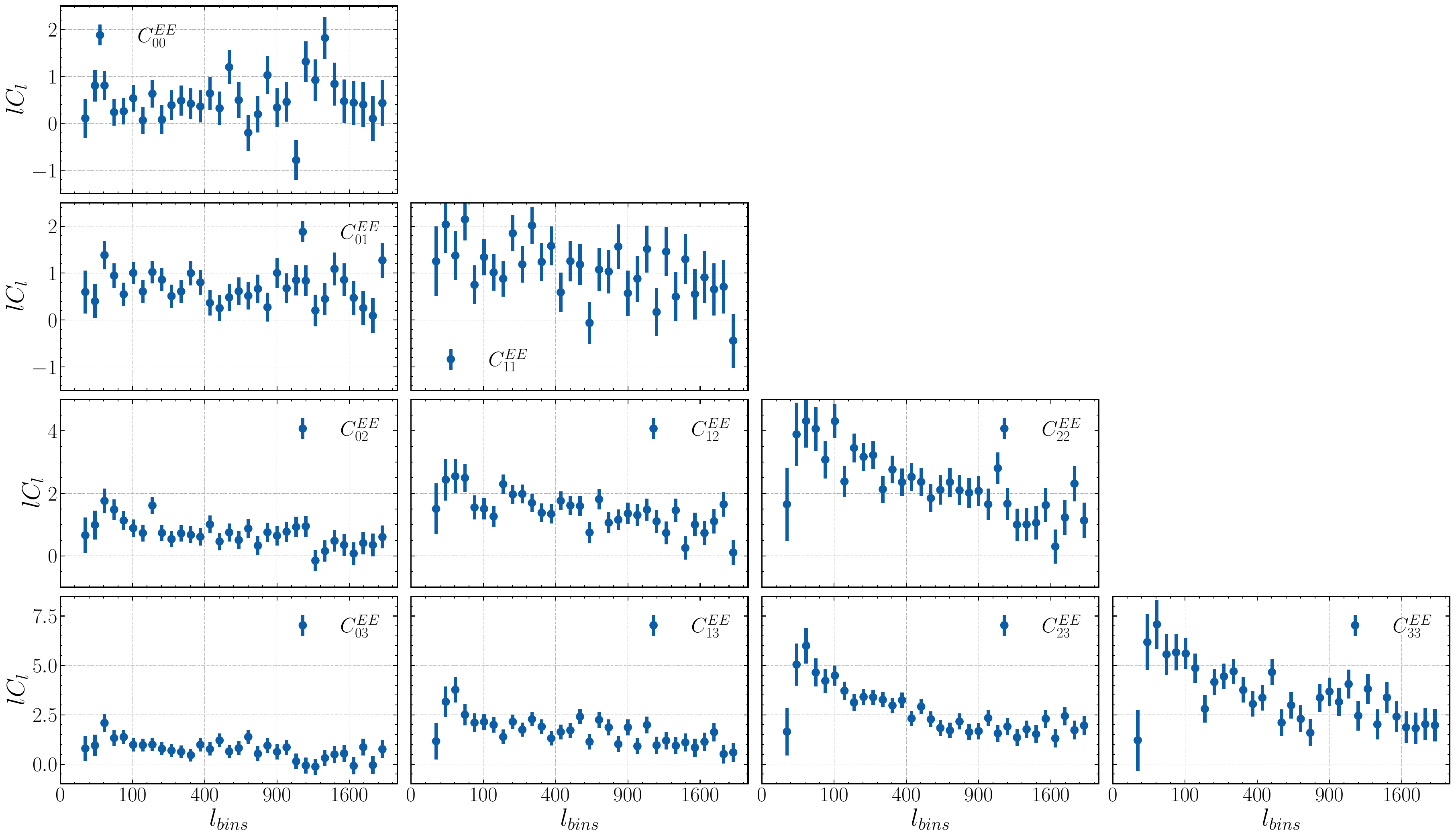}
    \caption{\emph{Top panel.} Shear ($E$-mode) power spectrum for DES Y3. The $C_l$'s are binned into 32 square-root spaced bins from $l_{min}=8$ to $l_{max}=2048$. The errors come from the diagonal covariance matrix computed analytically using \texttt{pymaster} under the DES-Y3 specifications.}
    \label{fig:shear_cell}
\end{figure*}

\begin{figure*}[ht!]
    \centering
    \subfigure{\includegraphics[width=0.96\textwidth]{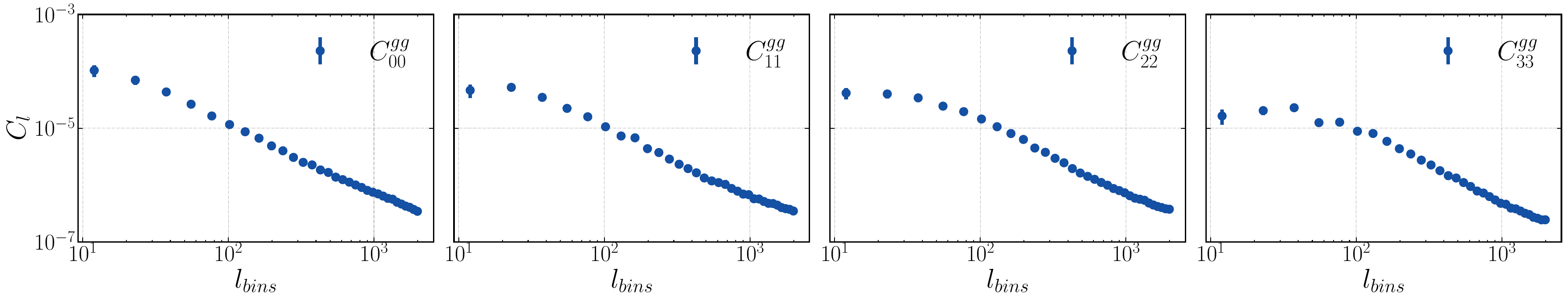}}
    \subfigure{\includegraphics[width=0.96\textwidth]{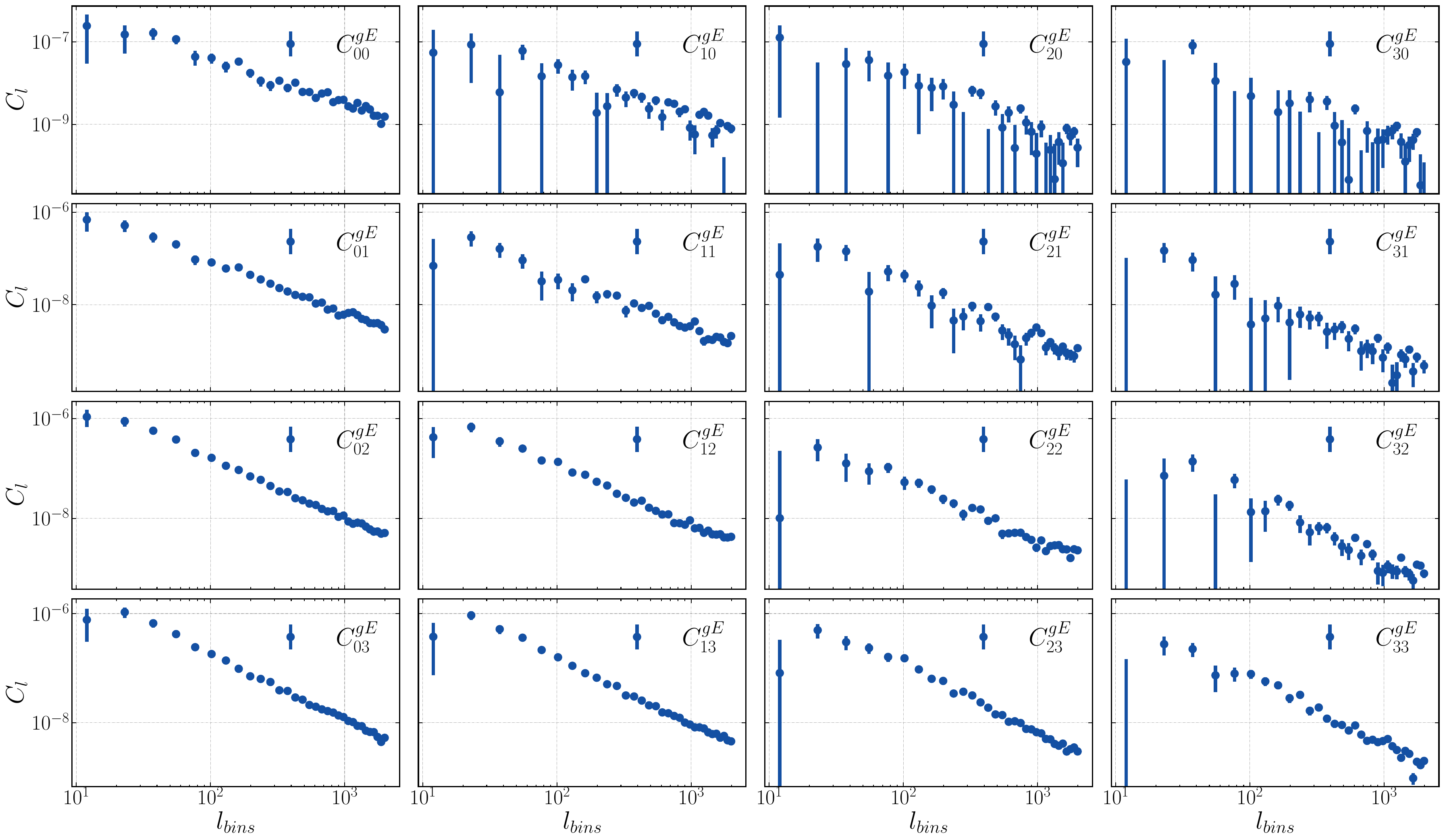}}
    \caption{\emph{Top panel.} Galaxy auto-spectrum $C_{gg}(l)$ for the four \maglim~ catalogue. Shot noise contribution has be removed. \emph{Bottom panel.} Galxaxy-galaxy lesning $C_{gE}(l)$ power spectra for the \maglim~ catalogue and \metacal~ source catalog for all tomographic bin combination.}
    \label{fig:galaxy_shear_results}
\end{figure*}

\subsection{$C_l$ estimation}

The angular power spectrum $C_l$ are defined for fields on full sky as
\begin{equation}
    \langle f^{a}_{l} {f^{b}_{l'}}^{\dagger}\rangle = C_l^{ab} \delta_{ll'}\delta_{mm'} 
\end{equation}
where $f^a$ and $f^b$ are scalar fields \footnote{While we describe the algorithm in terms of spin-0 fields, the formalism can be straightforwardly generalized to spin-2 fields.} defined on the full sky and $C_l^{ab}$ is the cross-spectrum between them.

Photometric surveys like DES survey partially sky and are thus sample masked version of full-sky cosmological fields. The masking in configuration space results in effective coupling of modes in harmonic space, making accurate power spectrum estimation and subsequent likelihood analysis, very challenging. In this work, we use the pseudo-$C_l$ framework implemented in the \texttt{pymaster} library \cite{Alonso:2018jzx} for estimation and likelihood analysis of angular power spectrum statistics of shear and galaxy fields. The pseudo-$C_l$ framework is a near-optimal approach for power spectrum estimation for masked photometric survey maps and we briefly describe the approach here and refer the readers to for more details.

A field $f$ in the sky mapped by a survey like DES with some complex masking/weighting $w$ can be expressed as
\begin{equation}
    \tilde{f^a}(\theta) = w^a(\theta)f^a(\theta)
\end{equation}
where $f^a$ is the true underlying cosmological field while $\tilde{f}^a$ is the masked version which we observe.
In harmonic space, we have
\begin{equation}
    \tilde{f}^a_l = \sum_{l'l''} D_{l'l''} w^a_{l'}f^{a}_{l''}
\end{equation}
where $D_{l'l''}$ is a spin-dependent coupling factor. As a result of this coupling, the cross-spectrum between fields $f^a$ and $f^b$ is given by
\begin{equation}
    C_l^{ab} = \sum_{l'} M_{ll'}C_l^{ab}
\end{equation}
where $M_{ll'}$ is the mode-coupling matrix for a given mask and leads to coupling/correlation between modes $l \ne l'$. The above relation is not easily invertible. The pseudo-$C_l$ algorithm instead first performs a binning operation on the coupled pseudo-$C_l$ and then employs a effective decoupling operation on the binned power spectrum to estimate the true, unbiased binned $C_L$ of the field.
\begin{equation}
    \hat{C_L^{ab}} = \sum (M^{ab})_{LL'}^{-1} \tilde{C}_{L'}^{ab}
\end{equation}
where 
\begin{equation}
    M^{ab}_{LL'} = \sum_{l \in L} \sum_{l' in L'}  M_{ll'}^{ab}
\end{equation}
where we have assumed that each $C_l^{ab}$ getting summed in a band appears with a constant weighting $w=1/N$ where $N$ is the number of multipoles $l$ contributing to the bandpower. Finally, we note that before comparing this estimate of bandpower $\hat{C_L^{ab}}$ to theory power spectrum in the likelihood analysis, one needs to \emph{forward} model the effect of binning and decoupling on the theory power spectrum.

\vspace{0.25cm}

The angular power spectrum estimation methodology described above is applied to DES Y3 maps of source and lens samples generated following the methodology described in \S\ref{sec:map_making}. For shear power spectrum estimation, we use the binning strategy from \cite{doux2022dark} to estimate the power spectrum of the $E$ mode of the shear field $\gamma$, $C_{EE}^{ij}(l)$ in 32 square-root spaced bins from $l_{min}=8$ to $l_{max}=2048$.  At linear-order, the $B$-mode power spectra is expected to be zero and therefore we exclude that from our analysis.



The galaxy-galaxy lensing power spectra $C_{gE}^{ij}$ is estimated for all the $4\times4=16$ bin combinations. For galaxy-galaxy clustering, we only estimate the auto-power spectrum $C_{gg}^{ii}(l)$ since cross-power spectrum is not expected to have much meaningful signal. We thus generate a total of 30 3x2 angular power spectrum combination each in 32 bins for $l=8$ to $l=2048$.

For the galaxy auto-spectrum, we estimate the noise contribution by first estimating the homogeneous Poisson noise from the observed galaxy number density and then applying the mask-dependent coupling operations using \texttt{pymaster} and subtract that from the total auto-spectrum. The lensing noise is similarly computed removed from the shear auto-spectrum. More details of the procedure can be found in \cite{nicola2021cosmic, doux2022dark, hadzhiyska2021hefty}. 

The estimated power spectra of DES Y3 maps are presented in \ref{fig:shear_cell} and \ref{fig:galaxy_shear_results}. 


\section{Modelling}
\label{sec:modelling}
We closely follow the prescription presented in \citep{DES:2021wwk} to model the 3x2 signal, noise and systematics. We work under the $\Lambda$CDM model with six cosmological parameters - $\Omega_m$, $\Omega_b$, $h$, $n_s$, $\sigma_8$ and $m_\nu$, which have their usual meaning. We use \texttt{CCL} public library \cite{ccl} with the default \texttt{CAMB} backend \cite{Lewis:1999bs} and use \texttt{`takahashi'} version of HALOFIT \cite{takahashi2012revising} to model non-linear power spectrum.

For our fiducial analysis where we use the linear bias model for galaxies, we have four linear bias parameters- $(b_1^1, b_1^2, b_1^3, b_1^4)$ for each of the four tomographic redshift bin. This choice of constant bias per bin has been found to be a very good approximation by \cite{DES:2021rex} and used in all of the DES Y3 analysis paper (however see \cite{Pandey:2023tjn} for a discussion of systematics associated with this simplified approach). The angular power spectra for galaxies in bin $i$ and $j$ is given by 
\begin{equation}
    C_{gg}^{ij}(l) = b_1^ib_1^jC(l)
\end{equation}
where $C(l)$ is the projected matter power spectrum for using Eq. ~\ref{limber_projection}. The bulk radial velocities of galaxies or the redshift space distortions produce negligible modification to the underlying power spectrum for DES Y3 hence we do not model them. Gravitation lensing of photons by intervening matter alters the number count of galaxies by altering the size and magnitude around the survey selection cutoff. We include a treatment for the resulting magnification bias using fixed bias provided by DES and given by $b_{mag} = (0.42, 0.30, 1.76, 1.94)$.

The observed ellipticities in the galaxy shapes are sourced not just by the gravitational shear but also by the large-scale tidal field in which the galaxies form and reside. This leads to correlated ellipticities which need to be modelled. To model this intrinsic alignment of galaxies, we use the non-linear alignment (NLA) model  \cite{2004PhRvD..70f3526H, Bridle_2007}. The bias in NLA model is given by
\begin{equation}
    b_{\rm IA}(z) = a\bar{C} \frac{\rho_{cric}\Omega_m}{D(z)}\bigg(\frac{1+z}{1+z_0}\bigg)^\eta
\end{equation}
where $a$ and $\eta$ are the parameters of the model, $z_0$ is the mean source catalogue redshift set to 0.62 in the analysis and $\bar{C}=5 \times 10^{-14} \mathrm{M_\odot} h^{-2} \mathrm{Mpc}^2$ is a normalization constant

A systematic bias to the redshift distribution $n(z)$ of the \maglim~ catalogue is modeled using  \emph{shift} and \emph{scaling/stretch} parameterizations i.e.\ there are four shift parameters $\mu_i$ to model a possible shift in the mean of $n^{i}(z)$ for each bin $i$ and four scaling parameters $\sigma_i$ which model possible dilation of the given distribution. The modeling can be mathematically expressed as
\begin{equation}
    n(z) \rightarrow \frac{1}{\sigma}n\bigg(\frac{z - \mu - \langle z \rangle}{\sigma} + \langle z \rangle\bigg) \label{eq:stretch}
\end{equation}
The dilation has been found to have a significant role in characterizing the lens sample properly. For the source sample, however, a four parameter modelling of a shift in the mean has been found to be sufficient in capturing any dominant systematic.

Finally, we model a possible multiplicative bias to the shear power spectrum via $m_{i}$ for each bins
\begin{equation}
    C^{ij}(l) = (1+m_i)(1+m_j)C^{ij}(l)
\end{equation}

This comprises our description of the 28-parameter fiducial model. A more detailed description could be found in \cite{DES:2021wwk}.

When performing our analysis using a nonlinear galaxy bias model, we will refer to the expression given in Eq.~(\ref{eq:Pgg}) for the galaxy one-loop power spectrum up to cubic order in bias operators. For a 3x2 analysis using the nonlinear galaxy bias model in Appendix \ref{sec:NL_model}, we include 8 additional free bias parameters, $b^i_2$ and $b^i_{\nabla \delta}$ for each of the 4 tomographic bins $i$. We set the remaining two nonlinear bias parameters as $b^i_{\mathcal{G}_2}=-2/7(b_1^i - 1)$ and $b^i_{\Gamma_3}=23/42(b_1^i - 1)$ using the parametric forms inspired from co-evolution models \cite{Chan:2012jj,Eggemeier:2018qae}. 

\textbf{Scale cuts:} In our fiducial 3x2 analysis, for the galaxy overdensity, we use simple but well motivated small-scale cut of $k_{\max}=0.1$ Mpc$^{-1}$ for each tomographic bin, in line with real-space cuts used in \cite{DES:2021rex, pandey2021dark}. For the shear component, we use the cuts generated in \cite{doux2022dark} for their fiducial analysis. Finally, after binning the data in square-root space, we discard the first three bins from each of the 3x2 statistics. This roughly corresponds to a large-scale cut of $l_{min}~45$. We discuss the motivation and reasoning behind these choices in  \S\ref{sec:discuss}.  The size of our data vector for the 3x2 analysis after these cuts is 177. Note however that for the 1x2 shear power spectrum analysis, we do not discard the low-l bins as we want to keep our analysis as close to \cite{doux2022dark} as possible. The size of the data vector for the 1x2 analysis is 119. In our analysis based on the nonlinear bias model, we consider a uniform scale cut of $k_{\max}=0.3$ Mpc$^{-1}$ for galaxy clustering and galaxy-galaxy lensing data vectors. We haven't fully explored the scale cuts for the nonlinear bias model and note that this particular choice is motivated by results presented in \cite{Nicola:2023hsd} where the authors obtain unbiased results from a similar nonlinear bias model up to $k_{\max} \lesssim 0.3$ Mpc$^{-1}$.

\section{Parameter Inference}
\label{sec:inference}
In order to infer the cosmological parameters using the 3x2 clustering data, we use the Bayesian framework. Under the Bayesian approach, the posterior distribution $\mathcal{P}(\Theta|D)$ over model parameters $\Theta$ of our fiducial model described in \S\ref{sec:modelling}, for a dataset $\mathcal{D}$ is given by
\begin{equation}
    \mathcal{P}(\Theta|D) \propto \mathcal{P}(D|\Theta)P(\Theta) = \mathcal{L}(D)P(\Theta) \\
\end{equation}
where $\mathcal{P}(\Theta)$ is the joint prior distribution over the model parameters and $\mathcal{L}(\Theta)$ is the likelihood. We assume prior independence and define $\mathcal{P}(\Theta)$ as a product of independent distributions; either uniform or Gaussian distributions. Our likelihood function is a Gaussian noise model for the band-powers given by
\begin{equation}
    \mathcal{L} = \exp(-\chi^2(\Theta, \mathcal{D})/2)
\end{equation}
with 
\begin{equation}
    \chi^2 =  -\sum_i \big(\mathcal{D}_i - C_i^{m}\big){Cov}^{-1} \big({\mathcal{D}_i - C_i^{m}}\big)
\end{equation}
where $\mathcal{D}$ is the data vector of 3x2 clustering band-powers estimated following the steps described in the last section and $C^{m}$ is the band-power estimated as a function of model parameters. The index $i$ runs over all the elements of the 3x2 band-powers left after scale cuts are applied. 
For the covariance $Cov$, we work under the Gaussian approximations and estimate disconnected contribution to the power spectrum covariance. We use \texttt{pymaster} to estimate the Gaussian covariance under the improved nearest kernel approximation (iNKA) following \cite{garcía-garcía2019disconnected}, \cite{nicola2021cosmic} and\cite{hadzhiyska2021hefty}. This has been shown to be a decent approximation for analysis in configuration space and also in the case of shear power spectrum analysis presented in \cite{doux2022dark}. The assumption of Gaussian likelihood for band-powers above $L\sim\mathcal{O}(50)$ is pretty good and has been shown to work well in the past. With our square-root spaced binning strategy, Gaussian likelihood assumption for $L<100$ remains an assumption which we will explore in a future work. 

We sample the posterior using \texttt{emcee} \cite{foreman2013emcee}. For our 28 parameter fiducial 3x2 analysis, we use the exact same priors used in \citep{DES:2021wwk} with the prior volume on $A_s$ linearly mapped to a volume on $\sigma_8$. The cosmic shear analysis uses a subset of the 28 parameters and the priors for those are also the same and come from \citep{DES:2021wwk, doux2022dark}.

\section{Results}
\label{sec:results}

\subsection{Cosmic shear analysis}

\begin{figure}[ht!]
    \centering
    \includegraphics[width=0.48\textwidth]{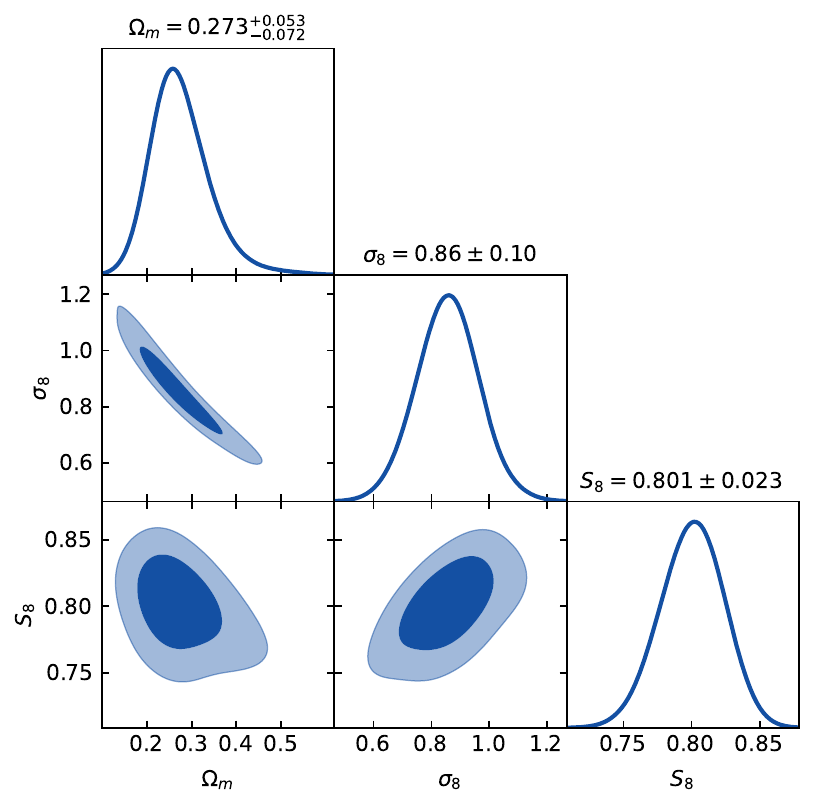}
    \caption{MCMC constraints from cosmic shear $E$-mode power spectrum analysis. We present 1D and 2D marginalized posterior distributions for $\Omega_m$, $\sigma_8$ and $S_8$. The contours in darker shade denote 68\% confidence interval with the lighter shade denoting the remaining region of the 95\% confidence interval.}
    \label{fig:1x2_results}
\end{figure}

We begin by presenting results for cosmic shear angular power spectrum analysis closely following the steps of \citet{doux2022dark}. Our dataset, including the data processing and preparations as well as the scale cuts, are kept as close as possible to the fiducial analysis of \citet{doux2022dark}. Our power spectrum covariance is a simple Gaussian covariance computed using iNKA approach at a fiducial cosmology using \texttt{CCL}. Our model includes 6 $\Lambda$CDM cosmology parameters together with four source distribution shift parameters $\mu^s_i$ and four calibration parameters $m^s_i$, one for each bin. Our intrinsic alignment model is a two parameter NLA model. Our results for the main parameters $\Omega_m, \sigma_8$ and $S_8\equiv\sigma_8 \sqrt{{\Omega_m}/{0.3}}$ are:
\begin{align*}
    \Omega_m = 0.273^{+0.053}_{-0.072} \\
    \sigma_8 = 0.86^{+0.10}_{-0.10} \\
    S_8 = 0.801^{+0.023}_{-0.023} 
\end{align*}
Our results are in excellent agreement with harmonic space analysis presented in  \citet{doux2022dark} and also with the configuration space analysis presented in \citet{DES:2021bvc, DES:2021vln}. The small $\sim0.3\sigma$ difference in $S_8$ compared to \citet{doux2022dark} possibly arises due to slight differences in our modelling choices. In Figure \ref{fig:1x2_results} we show the marginalized 1D and 2D posteriors for the relevant parameters. The posterior agrees well with Figure 12 of \citet{doux2022dark}. For the 119 element 1x2 data vector of angular power spectrum, our best-fit $\chi^2$ is 130.8 with p-value of $p=0.217$.

\subsection{3x2 analysis}
\begin{figure}[]
    \centering
    \includegraphics[scale=0.6]{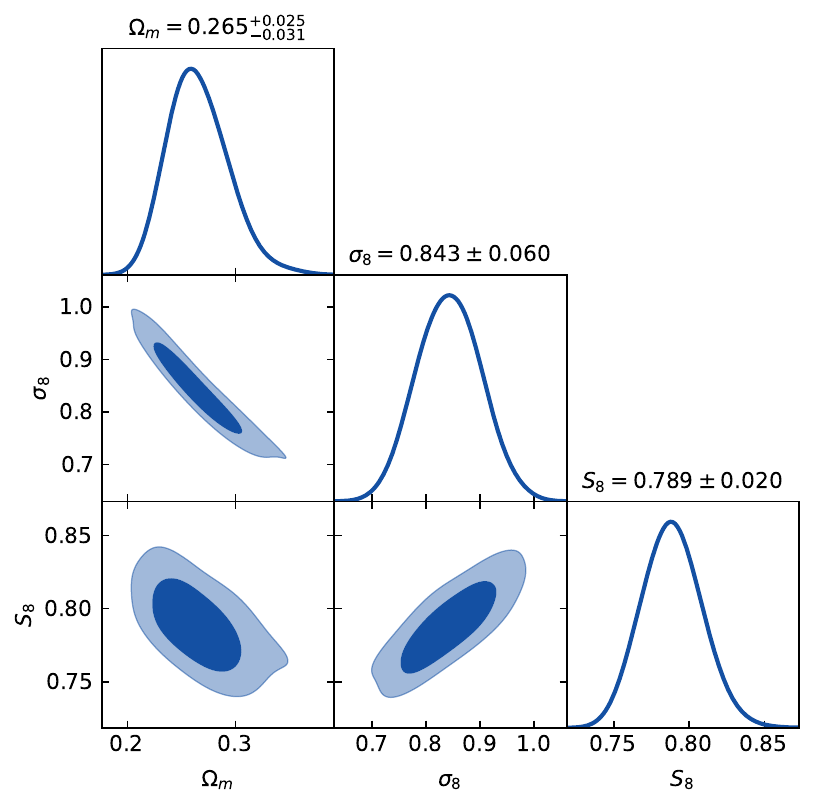}
    \caption{MCMC constraints from a linear-bias model based analysis of 3x2 clustering of DES Y3 data. We present 1D and 2D marginalized posterior distributions for $\Omega_m$, $\sigma_8$ and $S_8$. The contours in darker shade denote 68\% confidence interval with the lighter shade denoting the remaining region of the 95\% confidence interval.}
    \label{fig:3x2_results}
\end{figure}

In this section, we present the results of our 3x2 clustering analysis which uses 6-parameter $\Lambda$CDM model together with the linear model for galaxy bias, with an additional 18 nuisance parameters for including four lens distribution shift ($\mu^l_i$), four lens distribution dilation $\sigma^l_i$), four source distribution shift parameters $\mu^s_i$ and four calibration parameters $m^s_i$, one for each bin. We use a 2 paramater NLA model for intrinsic alignment. A detailed description of these modelling parameters is provided in \S\ref{sec:modelling}.

Our marginalized constraints for the main parameters $\Omega_m, \sigma_8$ and $S_8\equiv\sigma_8 \sqrt{{\Omega_m}/{0.3}}$ are:
\begin{align*}
    \Omega_m = 0.265^{+0.025}_{-0.031} \\
    \sigma_8 = 0.843^{+0.060}_{-0.060} \\
    S_8 = 0.789^{+0.020}_{-0.020} 
\end{align*}
In Figure \ref{fig:3x2_results}, we show the two-dimensional joint posterior distributions of these parameters as well as the marginalized 1D distributions. Our constraints broadly (within $\sim2\sigma$) agree with those presented in \cite{DES:2021wwk}.  In particular, the agreement for the $S_8$ values between the configuration space analysis of \citet{DES:2021wwk} and our harmonic space analysis is really good. 
For the 177 element 3x2 data vector of angular power spectrum, our best-fit $\chi^2$ is 174.8 with p-value of $p=0.533$. We reiterate that for our 3x2 analysis, we discard the first 3 bins of all our binned power spectra (corresponding to a $l_{min} \gtrsim 40$) but include those bins in our 1x2 analysis where we have tried to replicate the results of \cite{doux2022dark}. We discuss more differences between our analysis and that of \citep{DES:2021wwk} in the discussion section.

Our 3x2 analysis based on a nonlinear galaxy bias model consists of 30 nuisance parameters and 5 cosmological parameters. In this analysis, we set the energy density of massive neutrinos to zero. Our marginalized constraints are:
\begin{align*}
    \Omega_m = 0.306^{+0.026}_{-0.025} \\
    \sigma_8 = 0.781^{+0.035}_{-0.043} \\
    S_8 = 0.786^{+0.014}_{-0.016}.
\end{align*}Using 302 elements in our data vector for the 3x2 angular power spectra, we obtain a best-fit $\chi^2=346.2$  with a p-value of $p=0.041$. Our constraints agree with those presented in \cite{DES:2021wwk}.  Once again, we note a good agreement for the $S_8$ values between the harmonic space analysis in this work and the configuration space analysis of \cite{DES:2021wwk}.

\section{Discussion}
\label{sec:discuss}

Below we describe approximations made in our analysis and its limitations. Although we do not ourselves perform a simulation based validation of our approach, almost all our choices are guided and backed by validation performed by DES and other previous studies and we refer the reader to them. 
\begin{itemize}
    \item Our biggest assumption and limitation is the use of Gaussian covariance approximation which potentially underestimates error by $\mathcal{O}$(10)\%. Although \cite{friedrich2021dark} showed that non-Gaussian part of the covariance has a negligible impact on both maximum posterior and parameter constraints for a configuration space analysis, one would ideally want to show this explicitly for a harmonic-space analysis. This could be even more important for our nonlinear bias model where we consider data vectors up to a fiducial scale cut of $k_{\max}=0.3$ Mpc$^{-1}$. We postpone this to a later work along with a more detailed epxloration of the scale cuts for the nonlinear model.

    \item We bin modes of the angular power spectra in square-spaced bins following \cite{doux2022dark}. With our square-root spaced binning strategy, Gaussian likelihood assumption for $L\lesssim50$ remains untested that we will explore in a future work.

    \item When deciding on scale cuts for the 3x2 analysis, we follow a very conservative strategy and not only impose a small-scale cut for galaxies in $k_{max}=0.1 \mathrm{Mpc}^{-1}$  to mitigate issues related to bias and baryonic effects, but also a large-scale cut in $l_{min}$ for both galaxies and cosmic shear.  We exclude bins with $l\lesssim40$ corresponding to the first 3 bins under our square-space binning strategy for all the 3x2 clustering statistics. The \texttt{pymaster} based Gaussian covariance estimation is slightly inaccurate on $l<40$ thus discarding these modes is a good conservative choice. Additionally, the Gaussian likelihood model assumptions is a better description of the dataset sans the largest-scale modes. This also helps us evade any unknown systematics which typically afflict photometric surveys on the largest scales and produce spurious power. As a result, Limber approximation is good enough for our use case and so we work under Limber approximation for all our 3x2 clustering data.
    
    \item We use HALOFIT fitting formula to model linear and nonlinear matter power spectrum in our fiducial analysis. HALOFIT is fit on $N$-body simulation data and has an scale-dependent accuracy which can worse than 5\% at some scales of interest. Alternatives to HALOFIT include emulators like \textsc{cosmic emu} \cite{moran2022miratitan} \textsc{bacco} \cite{arico2020bacco} and \textsc{anzu} \cite{kokron2021cosmology} but these emulators have limited coverage in cosmological parameter space and hence can't be used to sample broad parameter space volume. \cite{DES:2021rex} found HALOFIT to be sufficiently accurate for DES Y3.
    \cite{friedrich2021dark}.

    \item We ignore B-modes in cosmic shear. The B-modes are not induced by gravitational lensing at linear order but can be induced by higher-order effects and by astrophysical effects. Moreover, $E/B$ mode mixing due to miscalibration of PSF is a systematic effect that can fake the B mode signal. Several diagnostic and validation studies have found B modes consistent with zero for DES Y3 \cite{gatti2021dark, doux2022dark}.
    
    \item We use the 2-parameter Non-linear Intrinsic Alignment model to characterize galaxy intrinsic alignments. This is a limiting case of the TATT model used in DES fiducial analysis but has been shown to be a sufficiently good model for DES data.
    
    \item The 1-sigma errors derived from the nonlinear bias model are noticeably reduced compared to our fiducial analysis. This reduction may, in part, be linked to the assumption of Gaussian covariance along with the access to a larger number of data points from smaller scales. Furthermore, it's worth emphasizing the importance of conducting a more comprehensive investigation into the scale cuts. We defer this to a future analysis.

\end{itemize}

Under these modeling assumptions and simplifications, we have analyzed \maglim~lens and \metacal~ source catalogues released by DES Y3, by modeling the two-point clustering of galaxy-galaxy, galaxy-lensing, and cosmic shear in harmonic space. Our results are consistent with the results of the same two-point clustering study performed by DES in configuration space, although we find a preference
for slightly lower $\Omega_m$ and a higher $\sigma_8$ in the harmonic space analysis. While we have tried to keep our analysis as similar to DES Y3 fiducial analysis as possible, differences remain that can explain this shift. We plan to use the pipeline developed in this work to investigate signatures of primordial non-Gaussianity in photometric datasets using its effect on large-scale galaxy power spectrum \cite{Dalal:2007cu} as well as cross-correlation of galaxy maps with other large-scale tracers \cite{Munchmeyer:2018eey, giri_ksz, schmittfull2018parameter}. Recently it was shown that non-perturbative approaches involving soft and collapsed limits of higher-order correlations \cite{giri_2023, Giri:2023mpg, Goldstein:2022hgr} can provide strong constraints on $f_{\rm NL}$ using very non-linear scales. It would be interesting to explore the viability of such an approach on photometric datasets.
Additionally, we are keen to investigate constraints on the amplitude of CDM blue-tilted isocurvature fluctuations \cite{Chung2021lfg}. As shown in \cite{Chung2023syw}, analysis of galaxy clustering for large blue-tilted isocurvature fluctuations requires a careful treatment of the divergences in various one and higher-order loop terms. This involves a consistent renormalization using relevant counter-terms and a complete operator basis up to cubic order in the bias expansion. It will be interesting to compare the constraints obtained from a 3x2 analysis with a recent forecast presented in \cite{Chung2023syw}.

\section*{Acknowledgements}
We thank Keith Bechtol and Moritz Muenchmeyer for useful discussions. Support for this research was provided by the University of Wisconsin - Madison Office of the Vice Chancellor for Research and Graduate Education with funding from the Wisconsin Alumni Research Foundation. We have extensively used several python libraries including \textsc{numpy}\cite{harris2020array}, \textsc{matplotlib}\cite{Hunter:2007}, \textsc{CLASS}\cite{Blas_2011}, \textsc{getdist}\cite{Lewis:2019xzd} and \textsc{SciencePlots}\cite{SciencePlots}. 

\appendix
\section{Nonlinear bias model \label{sec:NL_model}}

\subfile{nonlinear.tex}

\bibliography{refs}

\end{document}

%% file: preamble.tex
\usepackage{graphicx}
\usepackage{dcolumn}
\usepackage{bm}
\usepackage{float}
\usepackage[english]{babel}
\usepackage[utf8]{inputenc}
\usepackage{academicons}
\usepackage{subfigure}
\usepackage{dirtytalk}
\usepackage{amsthm}
\usepackage{mathtools}
\usepackage{graphicx}
\usepackage{float} 
\usepackage{physics}
\usepackage{xcolor}
\usepackage{adjustbox}
\usepackage{placeins}
\usepackage[T1]{fontenc}
\usepackage{subfigure}
\usepackage{verbatim}
\usepackage{csquotes}
\usepackage{mathalfa,amssymb}
\usepackage{xspace}
\usepackage[en-US]{datetime2}
\usepackage[pdftex, pdftitle={Article}, pdfauthor={Author}]{hyperref}

\newcommand{\orcid}[1]{\href{https://orcid.org/#1}{\textcolor[HTML]{A6CE39}{\aiOrcid}}}

\def\metacal{\textsc{metacalibration}}
\def\metacal{\textsc{des}}

\newcommand{\commentout}[1]{\ignorespaces}

\def\s8m{\texttt{s8\_m}\xspace}
\def\s8p{\texttt{s8\_p}\xspace}
\newcommand{\maglim}{{\texttt{Maglim}}}

\def\maglim{\texttt{MagLim}{\ignorespaces}}
\def\metacal{\textsc{Metacalibration}}

%% file: sai_file_1_theory.tex
\section{Theory}
\label{sec:theory}
In this section, we present a brief overview of the theoretical framework
underpinning the 3x2 analysis, which encompasses measurements derived
from cosmic shear, galaxy clustering, and galaxy-galaxy lensing. The
3x2 analysis, at its core, constitutes a statistical approach for
making cosmological inferences based on the two-point correlation
functions involving the observed projected galaxy density field, denoted
as $\delta_{g}$, and the weak lensing shear field, represented as
$\gamma$. These two fields' two-point auto and cross-correlations
yield three distinct sets of observables, as delineated below.

\subsection{Galaxy density field}
The observed galaxy over-density within a given tomographic bin $(i)$
when projected onto the celestial sphere can be expressed as a combination
of projected galaxy density contrast, modulation by magnification
($\mu$) and distortion from redshift-space measurements:
\begin{equation}
\delta_{g,{\rm obs}}^{i}(\vec{\theta})=\delta_{g,{\rm D}}^{i}(\vec{\theta})+\delta_{g,{\rm RSD}}^{i}(\vec{\theta})+\delta_{g,\mu}^{i}(\vec{\theta})
\end{equation}
where 
\begin{equation}
\delta_{g,{\rm D}}^{i}(\vec{\theta})=\int d\chi W_{g}^{i}(\chi)\delta_{g}^{i,{\rm (3D)}}\left(\vec{\theta}\chi,\chi\right)
\end{equation}
is the line-of-sight projection of the 3-D galaxy density contrast at a position $\vec{\theta}$ on the sky,
$\chi$ is the radial comoving distance to the redshift $z$ and 
\begin{equation}
W_{g}^{i}(\chi)=n_{g}^{i}(z)\frac{dz}{d\chi} \label{Eq:Wg}
\end{equation}
is the normalized window function of galaxies proportional to the
normalized number density distribution $n_{g}^{i}(z)$ of the lens
galaxy samples. We construct the 3D galaxy density contrast using
a pertrubative bias expansion consisting of all field level operators
allowed by Galilean symmetry 
\begin{equation}
\delta_{g}^{i,{\rm (3D)}}(x)=b_{1}^{i}\delta_{m}(x)+\frac{b_{2}^{i}}{2}\delta_{m}^{2}(x)+b_{\mathcal{G}_{2}}^{i}\mathcal{G}_{2}(x)+...
\end{equation}
where $\delta_{m}$ is matter overdensity field and $\mathcal{G}_{2}(x)$
is a second-order Galilean operator (See Appendix \ref{sec:NL_model}). Here we
have assumed that the bias values remain the same within a tomographic
bin.

The magnification term is given by
\begin{equation}
\delta_{g,\mu}^{i}(\vec{\theta})=C^{i}\kappa_{g}^{i}(\vec{\theta})
\end{equation}
with the magnification bias amplitude $C^{i}$, and where the tomographic
convergence field is given as
\begin{equation}
\kappa_{g}^{i}(\vec{\theta})=\int d\chi W_{\kappa,g}^{i}(\chi)\delta_{m}\left(\vec{\theta}\chi,\chi\right)\label{eq:convergence_field_kappa}
\end{equation}
with the lens efficiency window
\begin{equation}
W_{\kappa,g}^{i}(\chi)=\frac{3\Omega_{m}H_{0}^{2}}{2}\int_{\chi}^{\chi_{H}}d\chi'n_{g}^{i}(\chi')\frac{\chi}{a(\chi)}\frac{\chi'-\chi}{\chi'}.\label{eq:lens_efficiency_W}
\end{equation}
The RSD contribution $\delta_{g,{\rm RSD}}^{i}(\vec{\theta})$
is typically very small for photometric surveys and henceforth
we will not include it in our analysis.

%% file: nonlinear.tex
The galaxy clustering observable in harmonic space is determined from
the 3D galaxy-galaxy auto-correlation power spectrum $P_{gg}$. In
the linear bias theory, the galaxy/halo over-density is modeled at
linear order in underlying matter overdensity as
\begin{equation}
\delta_{g}=b_{1}\delta_{m}+\epsilon,
\end{equation}
where $\epsilon$ is the stochastic part which is not correlated with
the large-scale matter density field. The above linear model yields
the power spectrum $P_{gg}$ as
\begin{equation}
P_{gg}(k,z)=b_{1}^{2}(z)P_{\rm mm}(k,z)+P_{{\rm shot}}(z).
\end{equation}
In a purely perturbative approach, the matter power spectrum $P_{\rm mm}$
at linear bias order is taken to be the linear power spectrum $P_{{\rm lin}}$.
However, recent analyses (\cite{DES:2020yyz,Nicola:2023hsd}) have demonstrated that substituting the linear matter power spectrum $P_{mm}$ with the complete nonlinear power spectrum results in improved data fitting. This substitution also yields unbiased constraints, often accompanied by a modest reduction in the $\chi^2$ value, particularly from access to slightly smaller scales. For example, the DES Y3 fiducial 
analysis \cite{DES:2021wwk} uses the nonlinear matter power spectrum from the HALOFIT fitting function.\footnote{We note that HALOFIT does not include any treatment for baryonic effects.
These effects are important on scales $k\sim1$h/Mpc.} This approach extends the $k$-range over which one can probe the
observables with manageable theory error margins from a linear bias
expansion. The $k$-range can be determined by evaluating suitable scale-cuts such that one obtains unbiased results on the cosmological parameters. This is often achieved by testing the accuracy of the underlying theoretical model on mock galaxy catalogs.

However, it is well known that the linear bias theory is incomplete
since gravitational effects naturally induce higher-order operators. Hence, we also consider a nonlinear galaxy bias expansion (in Eulerian
coordinates ) including all operators allowed by Galilean
symmetry up to cubic order in the magnitude of the linear matter
overdensity $\delta^{(1)}$\citep{Desjacques:2016bnm}: 
\begin{align}
\delta_{g}(x) & =\sum_{\mathcal{O}}\left(b_{\mathcal{O}}+\epsilon_{\mathcal{O}}(x)\right)\mathcal{O}(x)+b_{\epsilon}\epsilon(x)\\
 & =b_{1}\delta(x)+b_{\epsilon}\epsilon(x)\nonumber \\
 & +\frac{b_{2}}{2}\delta^{2}(x)+b_{\mathcal{G}_{2}}\mathcal{G}_{2}(x)+\epsilon_{\delta}(x)\delta(x)\nonumber \\
 & +b_{\delta\mathcal{G}_{2}}\delta(x)\mathcal{G}_{2}(x)+\frac{b_{3}}{6}\delta^{3}(x)+b_{\mathcal{G}_{3}}\mathcal{G}_{3}(x)+b_{\Gamma_{3}}\Gamma_{3}(x) \nonumber \\
 & +\epsilon_{\delta^{2}}(x)\delta^{2}(x)+\epsilon_{\mathcal{G}_{2}}(x)\mathcal{G}_{2}(x)\nonumber \\
 & +b_{\nabla^{2}\delta}\nabla^{2}\delta(x)\label{eq:galaxy_bias_expn}
\end{align}
where all the operators $\mathcal{O}$ in the above expression are
considered to be coarse-grained and the subscript $\Lambda$ is dropped
for brevity. Eq.~(\ref{eq:galaxy_bias_expn}) is a double expansion in density
fluctuations and their derivatives since every insertion
of a Laplacian is equivalent to a second-order correction to an operator
$\mathcal{O}$.\footnote{In Fourier space the Laplacian takes the form $\nabla^{2}\rightarrow\left(k/k_{*}\right)^{2}$
where $k_{*}$ is some characteristic scale of clustering for biased
tracers and we restrict to scales $k/k_{*}\ll1$} Hence, the derivative operator in the last-line of Eq.~(\ref{eq:galaxy_bias_expn})
is counted approximately as cubic order in bias expansion. The operator sets $\{\delta^{2},\mathcal{G}_{2},\epsilon_{\delta}\delta\}$
and $\{\mathcal{G}_{2}\delta,\delta^{3},\mathcal{G}_{3},\Gamma_{3},\epsilon_{\delta^{2}}\delta^{2},\epsilon_{\mathcal{G}_{2}}\mathcal{G}_{2},\nabla^{2}\delta(x)\}$
are second and third order respectively and we refer the readers
to \citep{Desjacques:2016bnm} for details. Meanwhile, $\epsilon_{\mathcal{O}}$ in Eq.~(\ref{eq:galaxy_bias_expn})
are the stochastic noise contributions to galaxy formation. These are considered to be
uncorrelated with the long-wavelength fluctuations at
large scales.

The one-loop galaxy power spectrum at $O\left(\left(\delta^{(1)}\right)^{4}\right)$
can be written as 
\begin{align}
P_{gg}(k,z) & =b_{1}^{2}(z)P_{{\rm NL}}(k,z)+P_{gg}^{{\rm NLO}}(k,z)+P_{gg,\nabla^{2}\delta}(k,z) \nonumber \\
& +P_{gg,\epsilon}(k,z)\label{eq:Pgg}
\end{align}
where we take the non-linear matter power spectrum, $P_{{\rm NL}}$ from HALOFIT fitting function and evaluate the remaining next-to-leading order (NLO) one-loop contributions (without stochastic
terms) from \textsc{CLASSPT} \cite{Chudaykin:2020aoj}. The NLO one-loop power spectrum contributions are given as 
\begin{align}
P_{gg}^{{\rm NLO}}(k,z)/D^{4}(z)&=b_{1}(z)b_{2}(z)\mathcal{I}_{\delta^{(2)}\delta^{2}}(k) \nonumber \\
&+2b_{1}(z)b_{\mathcal{G}_{2}}(z)\mathcal{I}_{\delta^{(2)}\mathcal{G}_{2}}(k) \nonumber \\
& +b_{1}(z)\left(2b_{\mathcal{G}_{2}}(z)+\frac{4}{5}b_{\Gamma_{3}}(z)\right)\mathcal{F}_{\mathcal{G}_{2}}(k) \nonumber \\
 & +b_{2}(z)b_{\mathcal{G}_{2}}(z)\mathcal{I}_{\delta^{2}\mathcal{G}_{2}}(k)+\frac{1}{4}b_{2}^{2}(z)\mathcal{I}_{\delta^{2}\delta^{2}}(k) \nonumber \\
 &+b_{\mathcal{G}_{2}}^{2}(z)\mathcal{I}_{\mathcal{G}_{2}\mathcal{G}_{2}}(k)\label{eq:PNLO}
\end{align}
where $D(z)\equiv D_{+}(z)/D_{+}(0)$ is the normalized growth factor, and the loop contributions $\mathcal{I}_{\mathcal{O},\mathcal{O}'}(k)$
are given in \cite{Chung2023syw}. For the leading derivative term, we write
\begin{equation}
P_{gg,\nabla^{2}\delta}(k,z)=-2b_{1}(z)b_{\nabla^{2}\delta}(z)\left(\frac{k}{k_{*}}\right)^{2}P_{{\rm NL}}(k,z).\label{eq:P_lap}
\end{equation}
Our approach, which involves incorporating a nonlinear matter power spectrum for the leading higher-derivative bias terms, aligns with a similar strategy presented in previous works, including \cite{Nicola:2023hsd}. We set the clustering scale $k_*$ for each tomograhic bin $i$ to the following fiducial expression:
\begin{equation}
    k_*(z_i) = 0.4 ~D^{-4/3}(z_i) (\mbox{h/Mpc})
\end{equation}where $z_i$ is the mean redshift of tomographic bin $i$. This particular choice of redshift dependence for $k_*$ is motivated by a similar expression for $k_{\rm NL}$ obtained from linear dimensionless matter power spectrum. 
Contributions from other operators, such as $\delta^{3},\mathcal{G}_{3},\delta\mathcal{G}_{2}$
do not appear as they are eliminated during the renormalization process. In this analysis, we have ignored the stochastic contributions.